\newcommand{\CLASSINPUTtoptextmargin}{19mm}%
\newcommand{\CLASSINPUTbottomtextmargin}{43mm}%
\newcommand{\CLASSINPUTinnersidemargin}{12.9mm}%
\newcommand{\CLASSINPUToutersidemargin}{12.9mm}%
\documentclass[conference,10pt,a4paper]{IEEEtran}
\usepackage{amsmath}
\usepackage{amsmath,esvect}
\usepackage{times}
\usepackage{graphicx}
\usepackage{multirow}
\usepackage[none]{hyphenat}
\usepackage{float}
\usepackage{subfig}
\usepackage{iftex}
\usepackage[backend=bibtex,defernumbers=true,sorting=none,style=ieee]{biblatex}%
\ifPDFTeX%
\usepackage{t1enc}
\usepackage{times}
\fi%
\ifXeTeX%
\usepackage{fontspec}
\fi%
\ifLuaTeX%
\usepackage{fontspec}
\fi%
\makeatletter

\def\@maketitle{\newpage
\bgroup\par\addvspace{0.5\baselineskip}\centering%
\ifCLASSOPTIONtechnote
   {\bfseries\large\@IEEEcompsoconly{\sffamily}\@title\par}\vskip 1.3em{\lineskip .5em\@IEEEcompsoconly{\sffamily}\@author
   \@IEEEspecialpapernotice\par{\@IEEEcompsoconly{\vskip 1.5em\relax
   \@IEEEtitleabstractindextextbox{\@IEEEtitleabstractindextext}\par
   \hfill\@IEEEcompsocdiamondline\hfill\hbox{}\par}}}\relax
\else
   \vskip0.2em{\EuMWtitlesize\ifCLASSOPTIONtransmag\bfseries\LARGE\fi\@IEEEcompsoconly{\sffamily}\@IEEEcompsocconfonly{\normalfont\normalsize\vskip 2\@IEEEnormalsizeunitybaselineskip
   \bfseries\Large}\@title\par}\vskip1.0em\par
   \ifCLASSOPTIONconference%
      {\@IEEEspecialpapernotice\mbox{}\vskip\@IEEEauthorblockconfadjspace%
       \mbox{}\hfill\begin{@IEEEauthorhalign}\@author\end{@IEEEauthorhalign}\hfill\mbox{}\par}\relax
   \else
      \ifCLASSOPTIONpeerreviewca
         {\@IEEEcompsoconly{\sffamily}\@IEEEspecialpapernotice\mbox{}\vskip\@IEEEauthorblockconfadjspace%
          \mbox{}\hfill\begin{@IEEEauthorhalign}\@author\end{@IEEEauthorhalign}\hfill\mbox{}\par
          {\@IEEEcompsoconly{\vskip 1.5em\relax
           \@IEEEtitleabstractindextextbox{\@IEEEtitleabstractindextext}\par\hfill
           \@IEEEcompsocdiamondline\hfill\hbox{}\par}}}\relax
      \else
         \ifCLASSOPTIONtransmag
           {\@IEEEspecialpapernotice\mbox{}\vskip\@IEEEauthorblockconfadjspace%
            \mbox{}\hfill\begin{@IEEEauthorhalign}\@author\end{@IEEEauthorhalign}\hfill\mbox{}\par
           {\vspace{0.5\baselineskip}\relax\@IEEEtitleabstractindextextbox{\@IEEEtitleabstractindextext}\vspace{-1\baselineskip}\par}}\relax
         \else
           {\lineskip.5em\@IEEEcompsoconly{\sffamily}\sublargesize\@author\@IEEEspecialpapernotice\par
           {\@IEEEcompsoconly{\vskip 1.5em\relax
            \@IEEEtitleabstractindextextbox{\@IEEEtitleabstractindextext}\par\hfill
            \@IEEEcompsocdiamondline\hfill\hbox{}\par}}}\relax
         \fi
      \fi
   \fi
\fi\par\addvspace{0.0\baselineskip}\egroup}

\def\EuMWtitlesize{\@setfontsize{\EuMWtitlesize}{24}{24pt}}
\def\EuMWauthorsize{\@setfontsize{\EuMWauthorsize}{11}{11pt}}
\def\EuMWaffilsize{\@setfontsize{\EuMWaffilsize}{10}{10pt}}
\def\EuMWcaptionsize{\@setfontsize{\EuMWcaptionsize}{9}{10pt}}
\def\EuMWbibsize{\@setfontsize{\EuMWbibsize}{8}{10pt}}

\def\@IEEEauthorblockNstyle{\EuMWauthorsize\@IEEEcompsocnotconfonly{\sffamily}\@IEEEcompsocconfonly{\large}}
\def\@IEEEauthorblockAstyle{\EuMWaffilsize\@IEEEcompsocnotconfonly{\sffamily}\@IEEEcompsocconfonly{\itshape}\@IEEEcompsocconfonly{\large}}
\def\@IEEEauthordefaulttextstyle{\EuMWauthorsize\@IEEEcompsocnotconfonly{\sffamily}\sublargesize}

\def\thebibliography#1{\section*{\refname}%
    \addcontentsline{toc}{section}{\refname}%
    \EuMWbibsize\@IEEEcompsocconfonly{\small}\vskip 0.3\baselineskip plus 0.1\baselineskip minus 0.1\baselineskip
    \list{\@biblabel{\@arabic\c@enumiv}}%
    {\settowidth\labelwidth{\@biblabel{#1}}%
    \leftmargin\labelwidth
    \advance\leftmargin\labelsep\relax
    \itemsep \IEEEbibitemsep\relax
    \usecounter{enumiv}%
    \let\p@enumiv\@empty
    \renewcommand\theenumiv{\@arabic\c@enumiv}}%
    \let\@IEEElatexbibitem\bibitem%
    \def\bibitem{\@IEEEbibitemprefix\@IEEElatexbibitem}%
\def\newblock{\hskip .11em plus .33em minus .07em}%
\ifCLASSOPTIONtechnote\sloppy\clubpenalty4000\widowpenalty4000\interlinepenalty100%
\else\sloppy\clubpenalty4000\widowpenalty4000\interlinepenalty500\fi%
    \sfcode`\.=1000\relax}

\long\def\@makecaption#1#2{%
\ifx\@captype\@IEEEtablestring%
\par\@IEEEtabletopskipstrut
\else
\@IEEEfigurecaptionsepspace
\fi
\setbox\@tempboxa\hbox{\normalfont\footnotesize {#1.}\nobreakspace\nobreakspace #2}%
\ifdim \wd\@tempboxa >\hsize%
\setbox\@tempboxa\hbox{\normalfont\footnotesize {#1.}\nobreakspace\nobreakspace}%
\parbox[t]{\hsize}{\normalfont\footnotesize\noindent\unhbox\@tempboxa#2}%
\else
\ifCLASSOPTIONconference \hbox to\hsize{\normalfont\footnotesize\hfil\box\@tempboxa\hfil}%
\else \hbox to\hsize{\normalfont\footnotesize\box\@tempboxa\hfil}%
\fi\fi
\ifx\@captype\@IEEEtablestring%
\@IEEEtablecaptionsepspace
\else
\fi}

\newlength\tablecaptiontotableskip
\newlength\figuretocaptionskip
\def\@IEEEfigurecaptionsepspace{\vskip\figuretocaptionskip\relax}%
\def\@IEEEtablecaptionsepspace{\vskip\tablecaptiontotableskip\relax}%

\def\abstract{\normalfont%
\@IEEEabskeysecsize\bfseries\textit{\abstractname}\,\bfseries\textit{---}\,%
\@IEEEgobbleleadPARNLSP}%

\def\IEEEkeywords{\normalfont%
\@IEEEabskeysecsize\bfseries\textit{\IEEEkeywordsname}\,\bfseries\textit{---}\,%
\@IEEEgobbleleadPARNLSP}%
\def\endIEEEkeywords{\relax\vspace{0.67ex}%
\par\if@twocolumn\else\endquotation\fi%
\normalsize\normalfont}%

\DeclareRobustCommand*{\EuMWauthorrefmark}[1]{\raisebox{0pt}[0pt][0pt]{\textsuperscript{#1}}}%
\def\@IEEEauthorblockNtopspace{0ex}
\def\@IEEEauthorblockAtopspace{1mm}
\def\tablename{Table}
\def\thetable{\arabic{table}}
\def\IEEEkeywordsname{Keywords}
\def\subsubsection{\@startsection{subsubsection}{3}{\z@}{1.5ex plus 1.5ex minus 0.5ex}%
{0.7ex plus .5ex minus 0ex}{\normalfont\normalsize\itshape}}%
\newlength{\CPheadmatchindent}%
\def\@seccntformat#1{\hbox to\CPheadmatchindent{\csname the#1dis\endcsname}\hskip 0.1em \relax}
\IEEEilabelindentA \parindent
\IEEEilabelindent \IEEEilabelindentA
\IEEEelabelindent \parindent
\IEEEdlabelindent \parindent
\IEEElabelindent \parindent
\makeatother

\usepackage{amsmath}
\usepackage{times}
\usepackage{graphicx}
\usepackage{multirow}
\usepackage[none]{hyphenat}
\usepackage{float}
\usepackage{subfig}
\usepackage{iftex}

\usepackage{siunitx}
\usepackage{gensymb}
\usepackage{booktabs}
\usepackage[flushleft]{threeparttable} 
\usepackage{booktabs,caption}
\usepackage{xcolor}

\usepackage[
  separate-uncertainty = true,
  multi-part-units = repeat
]{siunitx}

\usepackage{manyfoot}%

\usepackage[acronym]{glossaries}
\usepackage{lineno}

\newcommand{\vcentered}[1]{\raisebox{-.5\height}{#1}}

\newacronym{RIS}{RIS}{reconfigurable intelligent surface}
\newacronym{DLA}{DLA}{delay line architecture}
\newacronym{RA}{RA}{resonant architecture}
\newacronym{UC}{UC}{unit cell}
\newacronym{PS}{PS}{phase shifter}
\newacronym{RE}{RE}{radiating element}
\newacronym{FoM}{FoM}{figure of merit}
\newacronym{NL}{NL}{normalized length}
\newacronym{RF}{RF}{radio frequency}
\newacronym{LC}{LC}{liquid crystal}
\newacronym{IMSL}{IMSL}{inverted microstrip line}
\newacronym{DGS-IMSL}{DGS-IMSL}{defected ground structure inverted microstrip line}
\newacronym{mm-Wave}{mm-Wave}{millimeter wave}
\newacronym{THz}{THz}{terahertz}
\newacronym{LCD}{LCD}{liquid crystal display}
\newacronym{PIN}{PIN}{positive intrinsic negative}
\newacronym{MEMS}{MEMS}{micro-electro-mechanical systems}
\newacronym{CMOS}{CMOS}{complementary metal-oxide-semiconductor}

\begin{document}
\raggedbottom
%
%
%


\title{Scalable mm-Wave Liquid Crystal Reconfigurable Intelligent Surfaces based on the\\ Delay Line Architecture}



%
%
\author{%
\IEEEauthorblockN{%
Julia Schwarzbeck\EuMWauthorrefmark{*},
Robin Neuder\EuMWauthorrefmark{*}, 
Marc Späth, 
Alejandro Jiménez-Sáez\EuMWauthorrefmark{$\dagger$}
}
\IEEEauthorblockA{%
Institute of Microwave Engineering and Photonics, Technische Universität Darmstadt, Darmstadt, Germany \\
\EuMWauthorrefmark{*} Robin Neuder and Julia Schwarzbeck contributed equally to this work\\
\EuMWauthorrefmark{$\dagger$}Corresponding author: alejandro.jimenez\_saez@tu-darmstadt.de  \\
}
}
\maketitle
%
%
%
\begin{abstract}
This paper presents the design, fabrication, and characterization of broadband liquid crystal~(LC) reconfigurable intelligent surfaces~(RIS) operating around 60\,GHz and scaling up to 750 radiating elements. 
The RISs employ a delay line architecture (DLA) that decouples the phase shifting and radiating layer, enabling wide bandwidth, continuous phase control exceeding 360°, and fast response times with a micrometer-thin LC layer of 4.6\,µm. 
Two prototypes with 120 and 750 elements are realized using identical unit cells and column-wise biasing. 
Measurements demonstrate beam steering over $\pm$60° and -3\,dB bandwidths exceeding 9\% for both apertures, confirming the scalability of the proposed architecture.
On top of a measured nanowatt power consumption per unit cell, aperture efficiencies above 20\% are predicted by simulations.
While the measured efficiencies are reduced to 9.2\% and 2.6\%, a detailed analysis verifies that this reduction can be attributed to technological challenges in a laboratory environment.
Finally, a comprehensive comparison between the applied DLA-based LC-RIS and a conventional approach highlights the superior potential of applied architecture.
 

\end{abstract}
%
%

\section{Introduction}
An earlier version of this paper was presented at EuMW 2025 and was published in its Proceedings \cite{neuder_a_broadband_750_2025}.

In the context of future communication standards, such as 6G and beyond, \glspl{RIS} have emerged as a solution for enabling reliable signal transmission with low latency and high data rates~\cite{ liu_reconfigurable_2021},\cite{bjornson_reconfigurable_2022}. These tunable surfaces promise to overcome challenges at \gls{mm-Wave} and \gls{THz} frequencies, addressing the issue of limited signal coverage due to high path loss, increased absorption, and reduced scattering.

\begin{figure}[t]
	\begin{center}
		\includegraphics[width=1\linewidth]{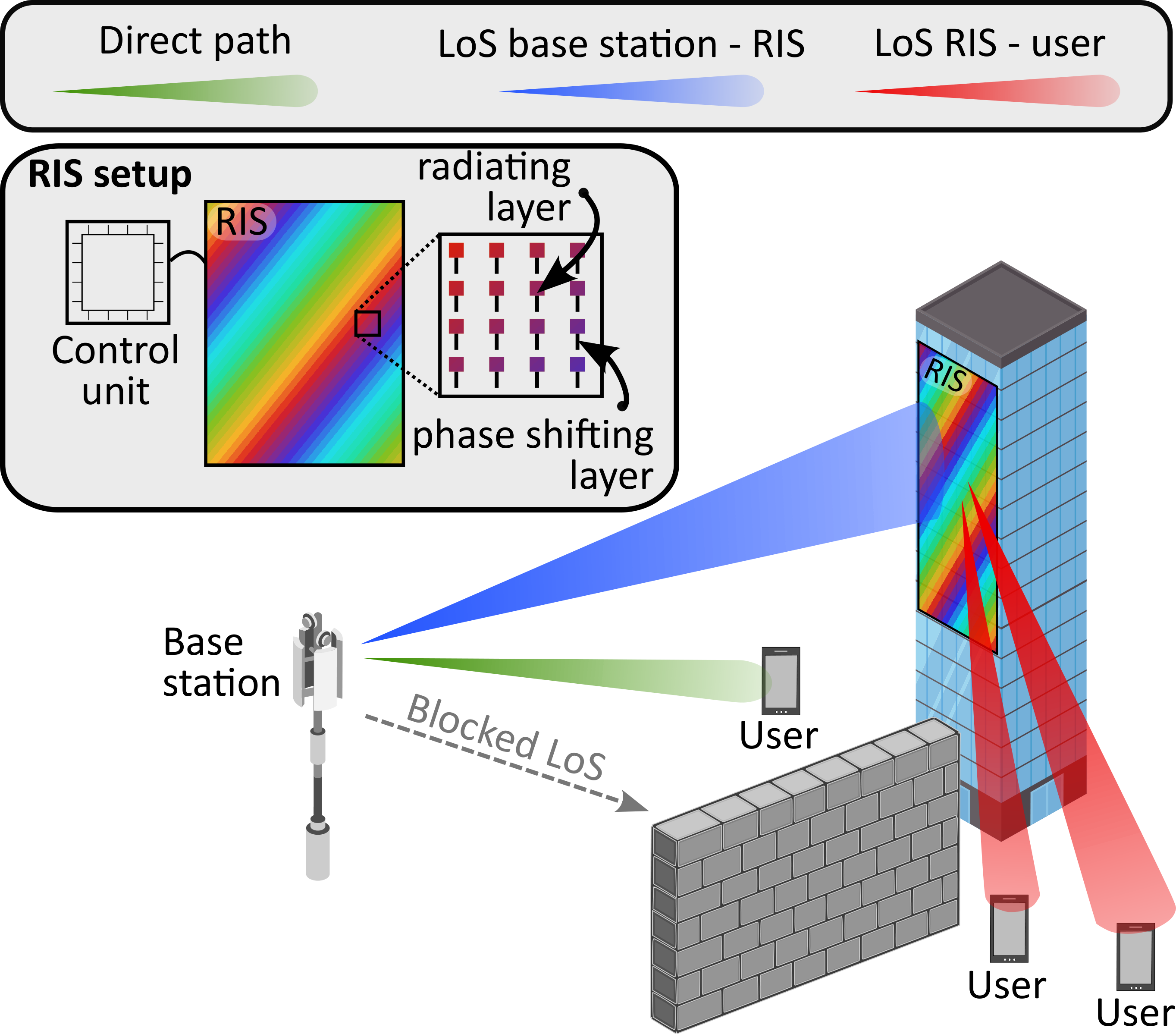}
		
		\caption{Exemplary outdoor scenario for a \gls*{RIS}~\cite{neuder_a_broadband_750_2025}. The different colors in the RIS qualitatively indicate different reflected phase at the radiating elements. LoS: Line of sight. RIS: Reconfigurable intelligent surface.
		}
		\label{Motivation}
		
	\end{center}
\end{figure}

\gls{RIS} can be envisioned as tunable reflectarrays, but without a defined feed.
They consist of an array of radiating elements arranged in a grid, along with a tuning mechanism that enables dynamic control of wave propagation.
In the far field, \gls{RIS} can redirect electromagnetic waves toward desired directions, while in the near field, they can also focus waves on focal points.
Fig.~\ref{Motivation} shows an outdoor scenario.
Large-scale \gls{RIS}, consisting of hundreds, thousands, or even millions of elements, are envisioned for future applications~\cite{bjornson_intelligent_2020},\cite{zhang_reconfigurable_2021}.
Most commonly, semiconductor-based tuning mechanisms, such as PIN~\cite{dai_reconfigurable_2020} and varactor diodes~\cite{pei_ris-aided_2021}, are used. 
While these technologies are readily available up to \gls{mm-Wave} frequencies, they come with drawbacks, particularly in terms of power consumption and cost: 
Given that at least one diode is required per radiating element, semiconductor-based \gls{RIS} quickly become prohibitively expensive as the array size increases~\cite{jimenez-saez_reconfigurable_2023}.

\begin{figure*}[ht!]
	\begin{center}
		\includegraphics[width=1\linewidth]{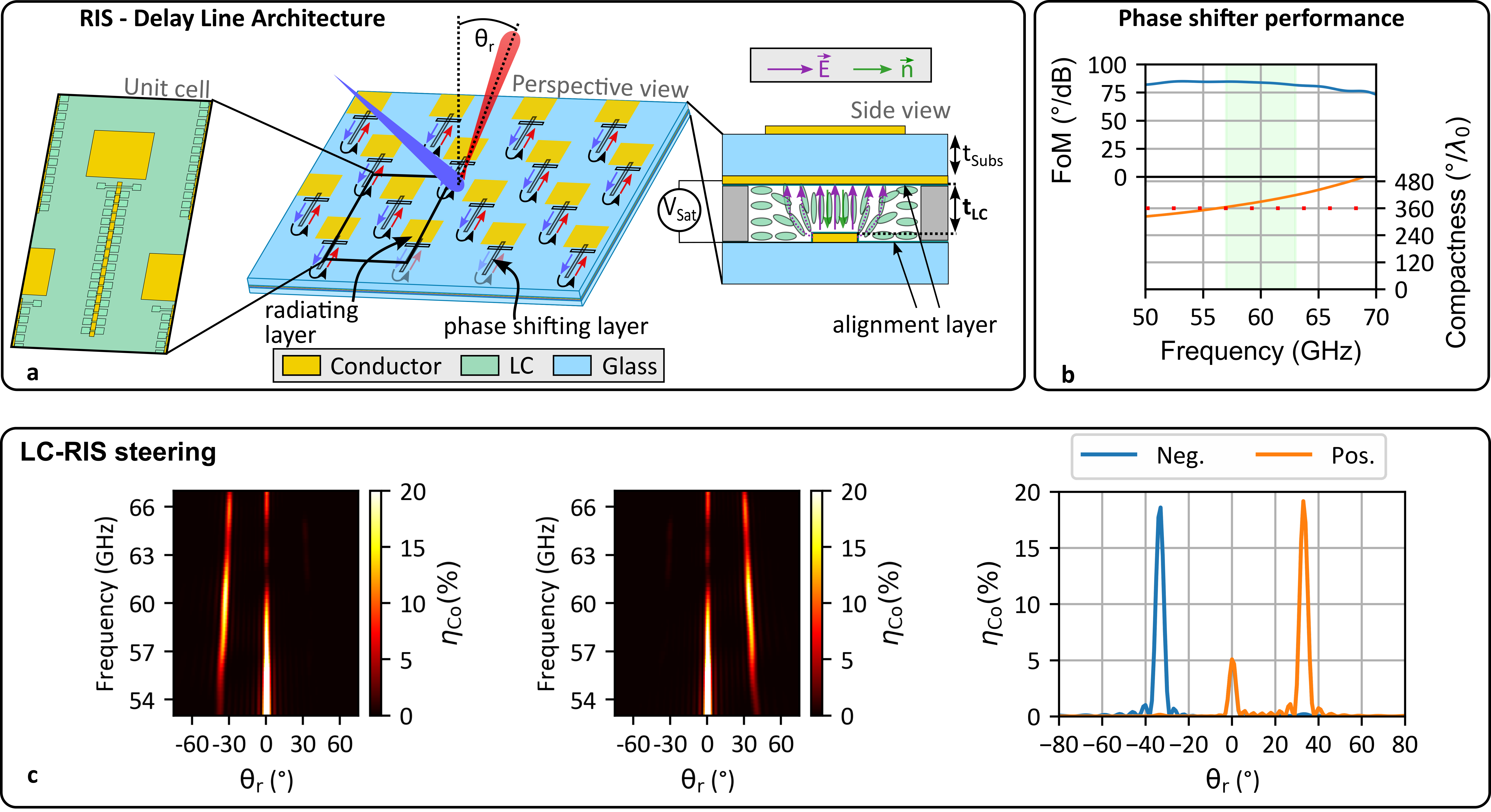}
		
		\caption{Design and simulation results of the proposed \gls{LC}-\gls{RIS}~\cite{neuder_a_broadband_750_2025}. a) Operation principle of the delay line architecture, the unit cell layout and the side view of the LC-RIS. b) Performance of the phase shifter in terms of losses~(FoM) and compactness. The red dotted line indicates a compactness of \SI{360}{\degree/\lambda_0}. c) Simulated aperture efficiency $\mathrm{\eta_{Co}}$ in a 768~($\mathrm{32 \times 24}$) element RIS. LC: Liquid crystal. RIS: Reconfigurable intelligent surface. FoM: Figure of Merit. $\theta_\mathrm{r}$:~reflected angle.
        }	
		\label{Simulation}
		
	\end{center}
\end{figure*}

A promising alternative tuning mechanism is nematic \gls{LC}~\cite{Jimenez_LC_goes_RF_2025}. 
Although it presents challenges such as moderate insertion losses, relatively slow response times, and temperature dependence, it also offers significant advantages. 
These include continuous tunability, transparency, low power consumption, and, most notably, large-scale scalability, making it a promising candidate for \gls{RIS} applications.
In~\cite{neuder_architecture_2024}, the \gls{DLA} was proposed for \gls{LC}-\gls{RIS}.
In contrast to conventional architectures relying on the detuning of a resonant element, called \gls{RA}, the \gls{DLA} couples the incoming electromagnetic wave to a tunable \gls{LC}-based phase shifter before retransmitting it into free space.
The \gls{DLA} allows \gls{LC}-\gls{RIS} to be optimized for low loss, fast response time, and wide bandwidth simultaneously. In~\cite{neuder_architecture_2024}, a \gls{DLA}-based \gls{LC}-\gls{RIS} operating around \SI{60}{\GHz} is presented.
The proposed \gls{LC}-\gls{RIS} is composed of 120~(12$\times$10)~tunable unit cell elements.

To demonstrate the scalability of the \gls{LC}-\gls{RIS} towards larger apertures, this paper extends the concept to an array with 750~(25$\times$30) elements, based on the same unit cell design.
The paper is structured as follows: First, the design of the proposed \gls{LC}-\gls{RIS} and its simulation results are presented.
Subsequently, measurement results of the manufactured \gls{LC}-\gls{RIS} prototypes are evaluated.
Thereby, the steering capabilities, the power consumption and the response time are analyzed.
Further, a measurement procedure for the aperture efficiency is proposed and the impact of fabrication tolerances are modeled.
Finally, the prototypes are compared to recent \gls{LC}-\gls{RIS} realizations to contextualize the potential of the \gls{DLA} in comparison to the conventional \gls{RA}.
\section{Design and simulation results}\label{sec:design_and_simulation_results}
The operating principle of the proposed LC-\gls{RIS} is illustrated in Fig.~\ref{Simulation}a).
The setup consists of three main materials: the \gls{LC} mixture, a glass substrate, and a conductor.
An electromagnetic wave propagating in free space impinges on the aperture-coupled patch antenna, which couples the wave to an LC-based phase shifter through the slot.
The wave propagates through the phase shifter to its open end, where the wave is reflected and subsequently re-radiated by the patch antenna.
The unit cell elements are arranged in a triangular grid to facilitate sufficient space for the phase shifters.
An element spacing $\le 0.5\lambda_0$ is targeted to fully avoid the appearance of grating lobes, with $\lambda_0$ being the free-space wavelength. 

The ellipsoidal shape of \gls{LC} molecules allows continuous permittivity tuning via bias voltages, which control the orientation of the molecule director $\mathrm{\vv{n}}$ relative to the \gls{RF} electric field $\mathrm{\vv{E}_{RF}}$~\cite{Jimenez_LC_goes_RF_2025}.
With the applied \gls{LC}-mixture GT7-29001 from Merck Electronics KGaA, Darmstadt, Germany, the permittivity can be tuned between $\mathrm{\varepsilon_{r,LC,\perp}}=2.46$ with $\mathrm{tan\delta_{LC,\perp}}=0.0116$ and $\mathrm{\varepsilon_{r,LC,\parallel}}=3.53$ with $\mathrm{tan\delta_{LC,\parallel}}=0.0064$~\cite{fritzsch_77-1_2019}.
For the glass substrate, AF32 from Schott is used~ ($\mathrm{\varepsilon_{r,AF32}}=5.1$ and $\mathrm{tan\delta_{AF32}}=0.009$ at \SI{24}{\GHz}~\cite{schott_electrical_2018}) and gold is employed as the conductor.

To maximize the phase shift per length and minimize the insertion losses, a \gls{DGS-IMSL} topology is employed.
The topology is well-suited for \gls{RIS} applications~\cite{neuder_compact_2023} which require low insertion loss, fast response time, high compactness, and wide bandwidth.
The simulated characteristics of the \gls{LC} phase shifter are presented in Fig.~\ref{Simulation}b).
The \gls{FoM} and the compactness are defined as:
\begin{equation}
\begin{alignedat}{4}
\mathrm{\gls{FoM}} &{}= \frac{\Delta \phi_\mathrm{max}}{\mathrm{IL_{max}}}
\qquad
\mathrm{Compactness} &{}= \frac{\Delta\phi_\mathrm{max}\, \lambda_0}{l_\mathrm{phys}}
\end{alignedat}
\end{equation}
where $\mathrm{\Delta \phi_\mathrm{max}}$, $\mathrm{IL_\mathrm{max}}$ and $l_\mathrm{phys}$ represent the phase shifter's max. differential phase shift, max. insertion loss and physical length, respectively.  
An \gls{FoM} of \SI{80}{\degree/dB} corresponds to \SI{4.5}{\dB} loss for $\SI{360}{\degree}$ phase shift and a compactness of $\SI{360}{\degree/\lambda_0}$ corresponds to a phase shift of $\SI{360}{\degree}$ differential phase shift per $\SI{0.5}{\lambda_0}$, as the wave travels back and forth through the phase shifter.

\begin{figure*}[t]
	\begin{center}
		\includegraphics[width=1\linewidth]{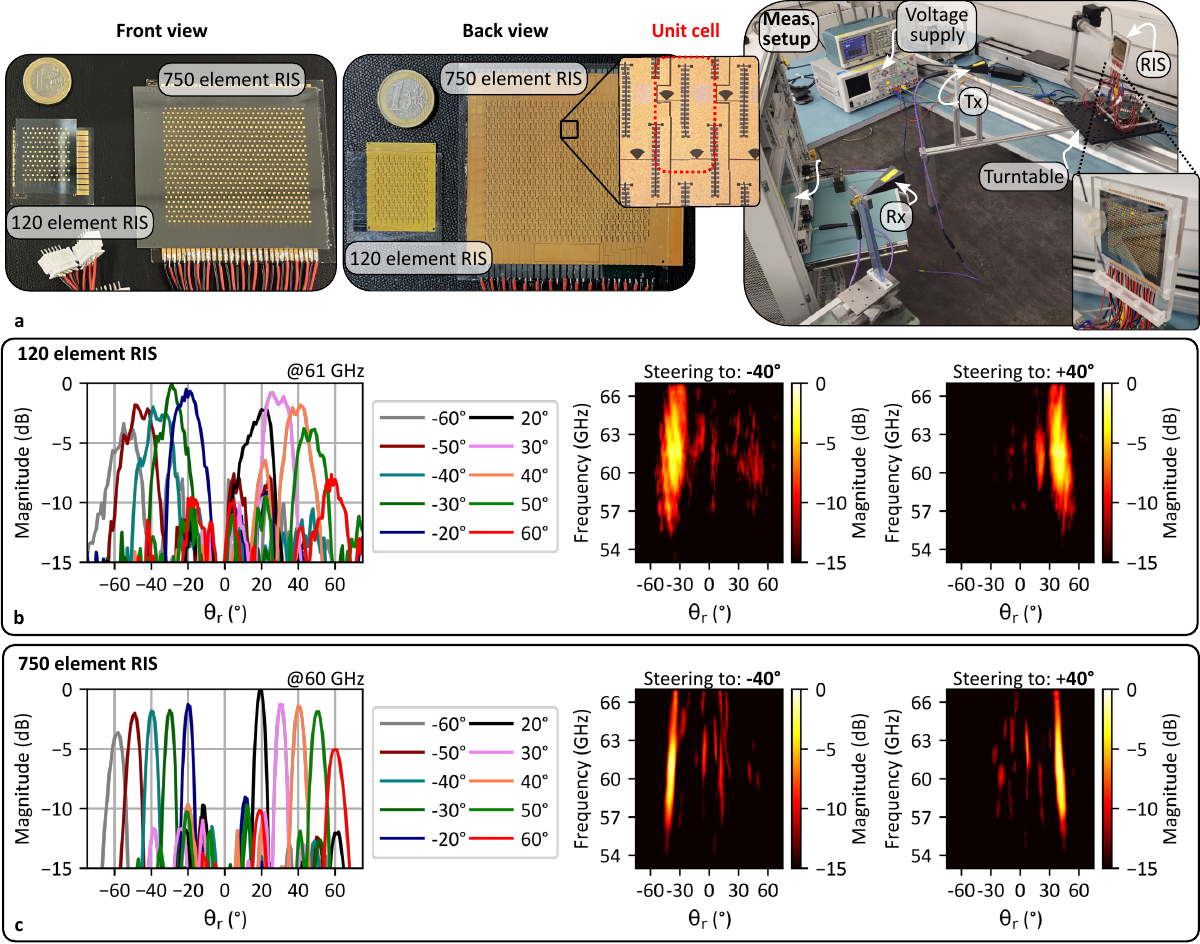}
		
		\caption{Measurement of the proposed \gls{LC}-\gls{RIS}~\cite{neuder_a_broadband_750_2025}. a) Front and back view of the \gls{LC}-\glspl{RIS} with 120 and 750 elements and the measurement setup. b)~Normalized measurement of the beam-steering capabilities of the 120 element \gls{RIS}. 
        c)~~Normalized measurement of the beam-steering capabilities of the 750 element \gls{RIS}. The dashed line in the right heat map indicates the impact of beam-squinting. \gls{RIS}: Reconfigurable intelligent surface. \gls{LC}: Liquid crystal. Tx: Transmitting antenna. Rx: Receiving antenna. $\theta_\mathrm{r}$: Reflected angle. VNA: Vector network analyzer.
        }
		
		\label{Measurement}
		
	\end{center}
\end{figure*}

The thickness of the \gls{LC} layer, $\mathrm{t_\mathrm{LC}}$ and the glass substrate are $\SI{4.6}{\mu \m}$ and $\SI{300}{\mu \m}$, respectively.
A thin \gls{LC} layer is essential, as the response times $\mathrm{\tau_{on}}$ (transition from $\mathrm{\varepsilon_{r,LC,\perp}}$ to $\mathrm{\varepsilon_{r,LC,\parallel}}$) and $\mathrm{\tau_{off}}$ (transition from $\mathrm{\varepsilon_{r,LC,\parallel}}$ to $\mathrm{\varepsilon_{r,LC,\perp}}$) increase with $\mathrm{t_\mathrm{LC}^2}$~\cite{jakoby_microwave_2020}.
To facilitate a faster transition of the molecule orientation back to its initial state after the bias is removed, a grooved alignment layer is applied to the surfaces.
Furthermore, an electrically thick glass substrate is required to achieve a large relative bandwidth.
The gold layers have a thickness of $\SI{2}{\mu \m}$.
The element spacing for the proposed \gls{LC}-\gls{RIS} is $\SI{0.45}{\lambda_0}$ within the triangular grid.

The simulated beam-steering capabilities of the proposed \gls{LC}-\gls{RIS} are demonstrated in Fig.~\ref{Simulation}c).
The \gls{RIS} is simulated comprising 768\footnote{Chosen close to 750 elements due to a simulation setup constraint: Multiple of an $\mathrm{4 \times2}$ cell for steering towards $\mathrm{\pm{33\degree}}$.}~($\mathrm{32 \times 24})$ unit cell elements.
The structure is excited by a plane wave impinging from broadside.
The simulation setup is described in more detail in~\cite{Neuder_DualPol_2025}.
The displayed results show the beam-steering in the H-plane, targeting angles around $\mathrm{\pm{33\degree}}$.
The aperture efficiency $\mathrm{\eta_{Co}}$ is evaluated relative to a perfectly conducting metal plate at the same distance, and peaks at 21.5\%.
The primary sources of loss originate from the lossy materials, contributing 28.6\%~(gold), 21.4\%~(\gls{LC}), and 18.2\%~(AF32)~at \SI{60.75}{\GHz}.
Outside its operating frequency, the \gls{LC}-\gls{RIS} behaves similar to a metal plate leading to a strong reflection to $\mathrm{\theta_r=0}$. 

\section{Measurement results}
Two \gls{LC}-\gls{RIS} are fabricated based on the unit cell above using standard lithographic processes~\cite{neuder_architecture_2024}.
SU-8 photoresist has been applied for the spacers between the two glass substrates constituting the \gls{LC} layer thickness, $t_\mathrm{LC}$.
The smaller prototype with 120~($\mathrm{10 \times 12}$) elements has already been analyzed in~\cite{neuder_architecture_2024}.
In addition, a larger prototype with 750~($\mathrm{30 \times 25}$) elements is manufactured.
Fig.~\ref{Measurement}a) displays both prototypes and the measurement setup.
See~\cite{neuder_architecture_2024} for a detailed description of the setup.
Reference measurements are taken and subtracted to isolate the \gls{RIS} from the cluttered environment.
In the following, the measured characteristics of the \gls{LC}-\gls{RIS} prototypes are analyzed.

\subsection{Steering Capabilities and Bandwidth}

The fabricated \gls{LC}-\glspl{RIS} support beam-steering in only one plane, due to a limitation to column-wise biasing.
Note that this limitation is purely technological, as element-wise biasing is already implemented in \gls{LC}-displays~\cite{neuder_architecture_2024},\cite{rose_active_2012},\cite{Widmaier_Process_2025}.
Beam-steering is achieved by applying \SI{1}{\kHz} square-wave bias voltages between 0 and $\mathrm{20\,V_{pp}}$.




In~Fig.~\ref{Measurement}b) and c), the measured steering capabilities of the 120 and 750 element \gls{LC}-\glspl{RIS} can be found.
The beam-steering of the large \gls{RIS} is comparable to the simulation results in Fig.~\ref{Simulation}c), but lacks the reflection to $\mathrm{\theta_r=0}$, as the latter is subtracted by the reference measurements. 
Both prototypes exhibit beam-steering capabilities between -60° and +60° and a \SI{-3}{dB} bandwidth between 9\%~(small prototype) and 9.5\%~(large prototype), proving the scalability of the design.
The bandwidth is evaluated corresponding to the reflected power at an exemplary angle $\theta_\mathrm{r}=40\degree$, corrected for beam-squinting to prevent underestimation in narrow beams.

\subsection{Power Consumption and Response Time}
Although not optimized for low power consumption, one \gls{LC}-\gls{RIS} element consumes only \SI{21.5}{\nW} on average, measured using an E4980A Precision LCR Meter.
Hence, for a \gls{RIS} with $\mathrm{10^6}$ elements (\SI{2.5}{m} x \SI{2.5}{m}
aperture with $\mathrm{\lambda_0/2}$ spacing at \SI{60}{\GHz}), the total power consumption would be \SI{21.5}{\mW}.
The response times of an \gls{LC} phase shifter with $t_\mathrm{LC}=\SI{4.6}{\mu \m}$ is measured to $\tau_\mathrm{on} = \SI{15}{\ms}$ and $\tau_\mathrm{off} = \SI{72}{\ms}$~\cite{neuder_compact_2023}.
The measured total response time of the larger LC-RIS prototype is $\tau_{\mathrm{RIS}}\approx\SI{250}{\ms}$, which is higher than expected.
The undesired increase most probably arises from a non-ideal structuring of the alignment layer~\cite{neuder_architecture_2024}.
By solving these technological imperfections, response times of only few tens of milliseconds may be attained~\cite{jakoby_microwave_2020}.

\subsection{Aperture Efficiency}
The aperture efficiency $\eta_\mathrm{Co}$ represents how effectively the physical aperture of the RIS contributes to the desired backscattered wave, both in terms of steering direction and polarization purity.
Losses occur as portions of the incident power are dissipated as heat within the device materials (glass, gold, and LC) or excite scattering modes that lead to differing polarizations or radiation into undesired directions.
The aperture efficiency can be defined relative to the ideal reflection from a perfectly conducting metal plate of the same aperture size, for which the radar cross section (RCS) can be determined analytically:
\begin{equation}
    \mathrm{\sigma_{MP} = \frac{4\pi A^2_{MP}cos(\theta_{Tx})cos(\theta_{Rx})cos(\phi_{Tx})cos(\phi_{Rx})}{\lambda_0^2}} ,\label{eq:sigma_MP}
\end{equation}
where $\mathrm{A_{MP}}$ corresponds to the aperture size of the metal plate and $\mathrm{\theta_{Rx/Tx}}$ and $\mathrm{\phi_{Rx/Tx}}$ correspond to the angles of incoming and reflected waves with respect to the metal plate.
Based on the measurement of the received powers for the RIS and the metal plate, i.e. the scattering parameters $\mathrm{S_{21,RIS,dB}}$ and $\mathrm{S_{21,MP,dB}}$, and the analytically calculated RCS of the metal plate, the RIS RCS $\mathrm{\sigma_{RIS}}$ is calculated according to
\begin{equation}
    \mathrm{\sigma_{RIS} = \sigma_{MP} + S_{21,RIS,dB} - S_{21,MP,dB}}.
\end{equation}
With this result, the aperture efficiency of the RIS is given by
\begin{equation}
    \mathrm{\eta_{Co,meas} = \frac{\lambda_0^2 \sigma_{RIS}}{4\pi A_{RIS}^2cos(\theta_{Tx})cos(\theta_{Rx})cos(\phi_{Tx})cos(\phi_{Rx})}}, \label{eq:eta_co_meas}
\end{equation}
where $\mathrm{\theta_{Rx}}$, $\mathrm{\theta_{Tx}}$, $\mathrm{\phi_{Rx}}$ and $\mathrm{\phi_{Tx}}$ are the incoming and reflected angles in azimuth and elevation with respect to the RIS.
$\mathrm{A_{RIS}}$ is the aperture size of the RIS, calculated as
\begin{equation}
    \mathrm{A_{RIS} = d_x \, d_y \, M \, N},
\end{equation}
with $\mathrm{d_x}$ and $\mathrm{d_y}$ being the element spacings and M and N the number of rows and columns.
Theoretically, a metal plate with an arbitrary aperture size can be taken for evaluation since the aperture of the surface is included in Eq.\,\ref{eq:sigma_MP}.
However, using a metal plate with dimensions matching those of the RIS is preferable, as it helps minimize the effects caused by variations in the illuminated areas of the horn antennas' beam widths, i.e., the same illumination efficiency.
Just like the RIS measurements, the metal plates are evaluated with subtraction of a reference measurement to isolate their response from the environment.
The measurement setup for the aperture efficiency can be seen in Fig. 1 of the supplementary material.

\begin{figure}[h]
	\begin{center}
		\includegraphics[width=1\linewidth]{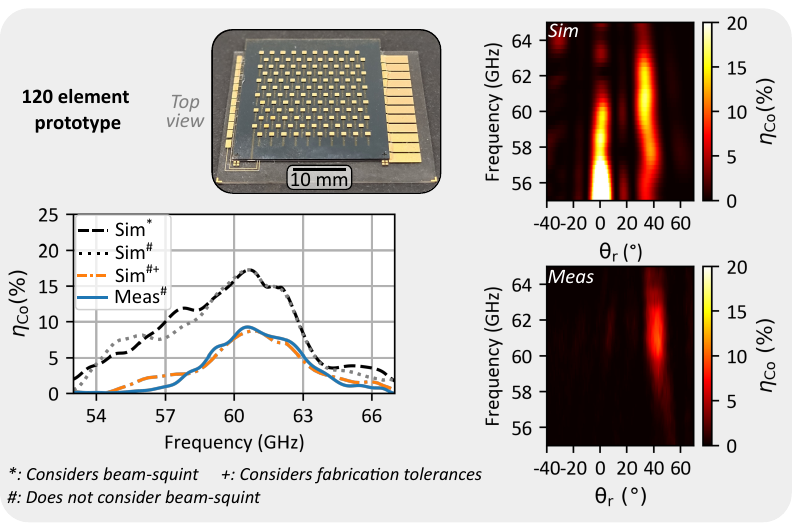}
		\caption{Aperture efficiency results of the RIS prototype with 120 elements.}
		\label{fig:Apt_eff_120}
	\end{center}
\end{figure}

The aperture efficiency of the small RIS prototype is illustrated in Fig. \ref{fig:Apt_eff_120}.
It features a maximum simulated aperture efficiency of 17.2\,\%.
This value is lower than the 21.5\,\% found by the simulation results from section \ref{sec:design_and_simulation_results}, arising from the additional consideration of the RF choke compared to the ideal model used before.
The effect of beam-squinting on the reflection angle $\mathrm{\theta_r}$ is almost negligible due to the small aperture size as well as the rather small steering angle, which is confirmed by simulations.
The measured aperture efficiency (blue) with a peak of 9.2\,\% is considerably lower than the simulation results.
Additionally, the measured bandwidth shows a reduction with respect to the simulations. 
The missing specular reflection in measurements also becomes evident, which is a side-effect of the reference measurement subtraction as explained in \cite{neuder_architecture_2024}.

\begin{figure}[h]
	\begin{center}
		\includegraphics[width=1\linewidth]{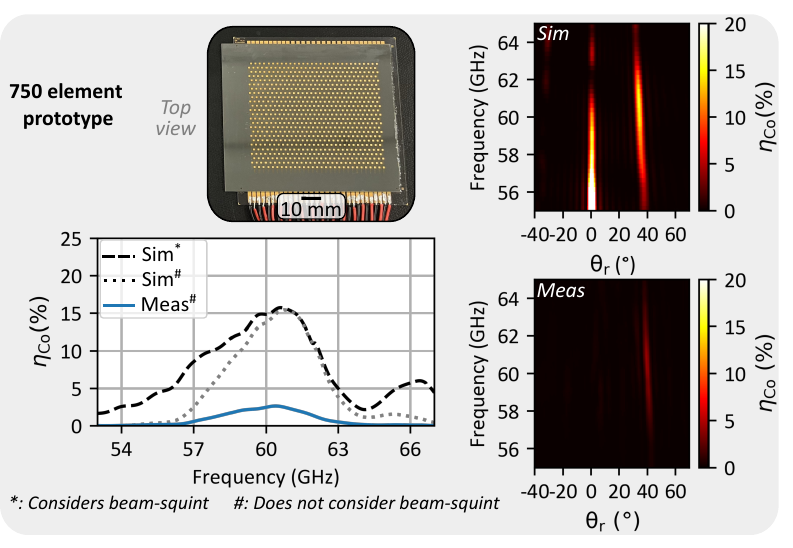}
		\caption{Aperture efficiency results of the RIS prototype with 750 elements.}
		\label{fig:Apt_eff_750}
	\end{center}
\end{figure}
The outcome for the prototype with 750 elements is illustrated in Fig. \ref{fig:Apt_eff_750}.
The results reveal a much lower measured aperture efficiency with a maximum of approximately 2.6\,\%. 
The maximum aperture efficiency obtained in simulation is 15.7\,\%.
Due to the larger aperture size, the results show a comparatively more pronounced effect of beam-squinting, both in simulation and measurement.


\subsection{Impact of Fabrication Tolerances}
Regarding the considerable reduction of the aperture efficiency and the smaller bandwidths, two possible causes may be considered: i) increased material loss and
ii) fabrication tolerances.
The impact of i) is presumably rather small, since 
the same materials showed a good agreement between simulation and measurement~(in~\cite{neuder_architecture_2024} and \cite{neuder_compact_2023}).
Hence, ii) fabrication tolerances are assumed to have a major impact on the measured performance.
Particularly, an uneven \gls{LC} thickness (see Fig.~\ref{Simulation} a)) and misalignments between the metallization layers are supposed to affect the measured outcome.

\begin{figure}[h]
	\begin{center}
		\includegraphics[width=1\linewidth]{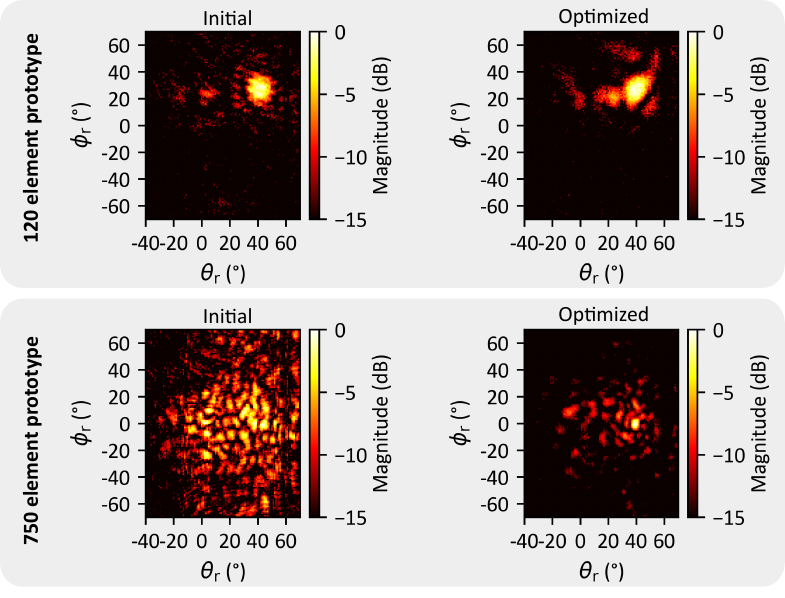}
		\caption{Measured distribution of back-scattered power of the RIS prototypes over azimuth and elevation for an initial voltage distribution and an optimized voltage distribution. Each graph is normalized to its respective maximum.}
		\label{fig:back-scattered_power}
	\end{center}
\end{figure}

To verify this assumption, first, the distribution of the back-scattered power in azimuth and elevation is examined for both prototypes in Fig.~\ref{fig:back-scattered_power}. 
The aim is to analyze whether the \gls{LC}-\glspl{RIS} produce a focused beam (as desired) or whether they scatter the incoming energy to various directions.
Within this context, an initial bias voltage set, which considers a uniform $t_\mathrm{LC} = \SI{4.6}{\micro \m}$ layer is compared to a bias voltage set that is optimized towards maximum received power~\cite{neuder_architecture_2024}. 
Through an optimization of the bias voltages for the desired steering angle, fabrication tolerances, such as an uneven $t_\mathrm{LC}$ across the surface, can be compensated to a certain extent, improving the focusing capabilities of the RIS.

\begin{figure}[h]
	\begin{center}
		\includegraphics[width=0.875\linewidth]{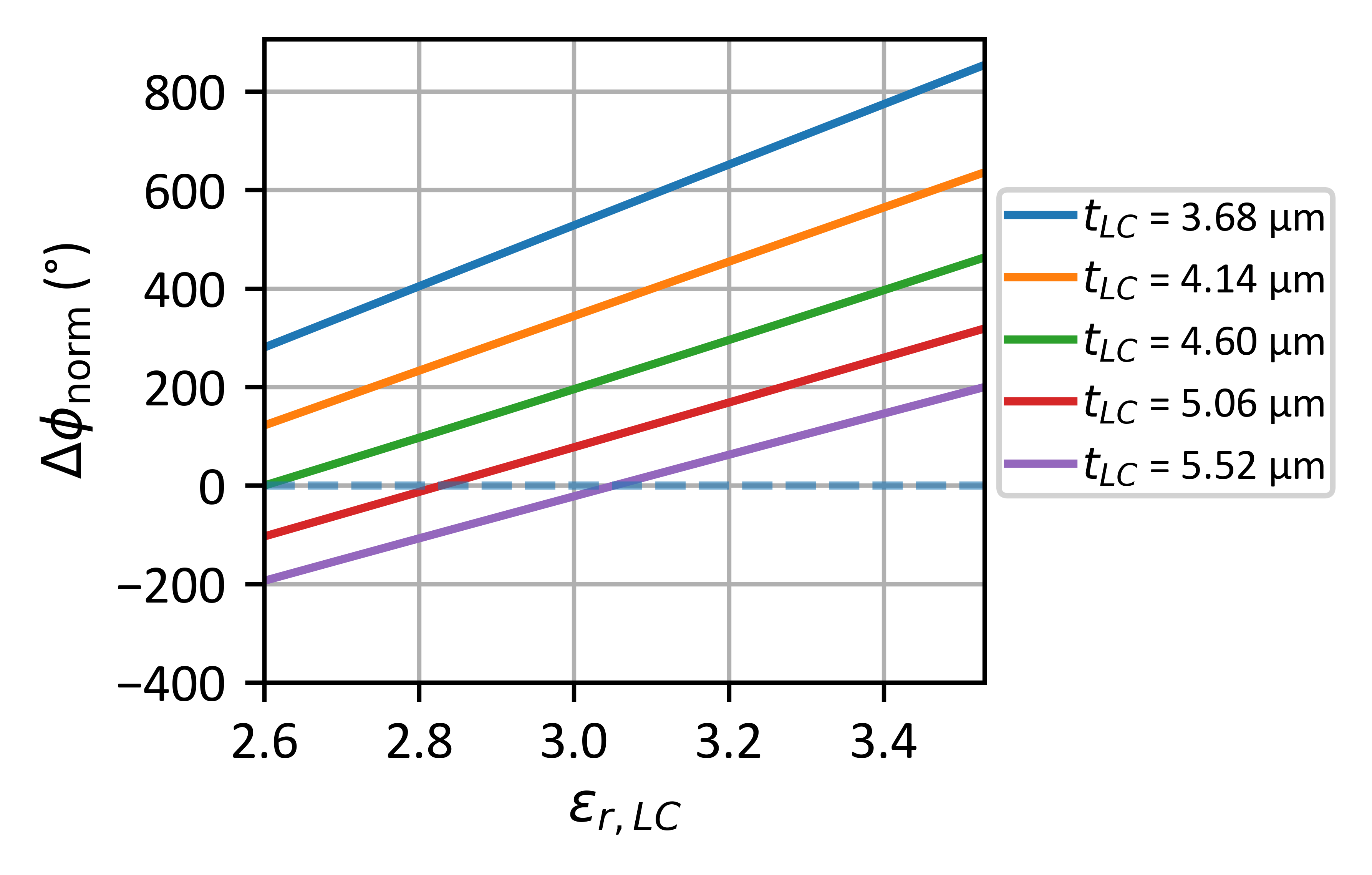}
		\caption{Influence of varying $t_\mathrm{LC}$~(see Fig.~\ref{Simulation}) on the inter-element phase shift $\Delta \phi$. 
        The phase shifts are presented with regard to the unbiased case ($\mathrm{\varepsilon_r}=2.6$) with thickness $t_\mathrm{LC}=\SI{4.6}{\micro \m}$.
        }
		\label{fig:phases_for_t_LC_variations}
	\end{center}
\end{figure}

Considering the smaller prototype, the reflected beam seems to be well focused of the reflection for the set of initial bias voltages (upper left subplot).
However, angular offsets are present, as the device was intended to steer toward $\theta_\mathrm{r} = 30\degree$ (azimuth), but shows its maximum at $\approx42\degree$.
Furthermore, since the elevation plane does not offer steering capabilities, the beam would ideally steer towards $\phi_r = 0\degree$, but is found at $28\degree$.
Since the offsets are consistent for all steering angles, they are expected to arise from the impact of a non-uniform, tilted \gls{LC} thickness that increases towards one corner of the surface.
The simulated effect of varying $t_\mathrm{LC}$ on the resulting phase of a unit cell is shown in Fig.~\ref{fig:phases_for_t_LC_variations}.
In essence, the plot shows that, given a fixed \gls{LC} permittivity, the differential phase shift is significantly affected by even small variations in \gls{LC} thickness.
The impact of this is studied in the supplementary material in more detail, where a model shows, that the tilted \gls{LC} surface reduces the maximum magnitude by $\approx \SI{3}{\dB}$.
Hence, the uneven \gls{LC} thickness is assumed to be the major reason for the overall decreased aperture efficiency.
For the small prototype, the voltage optimization barely increases the maximum received magnitude by $\approx \SI{0.3}{\dB}$.

Similar to the lower aperture efficiency, the decreased bandwidth of the small prototype is most likely due to fabrication imperfections, especially due to a slightly misaligned upper and lower glass substrate. 
Simulations show that a misalignment~($\le30\,\mu m$) between the metallization~(ground plane, phase shifter and antennas) does not have a decisive influence on the maximum aperture efficiency, but rather on the bandwidth.
Together with the tilted \gls{LC} layer thickness, this effect is considered in the orange curve in Fig.~\ref{fig:Apt_eff_120}, which agrees well with the measurement results. 
It should be noted that the model, which has been adjusted to align with the measurement results, is intended to illustrate the potential impact of fabrication tolerances rather than provide an accurate representation of the fabricated prototype.


The lower left subplot of Fig.~\ref{fig:back-scattered_power} shows the initial back-scattered power of the larger 750 element prototype with the bias voltages set to steer toward $\theta_r=40\degree$.
No distinct main beam is present, but the wave is scattered to various directions without a clear focus.
The  behavior is assumed to also result from a non-uniform $t_\mathrm{LC}$, which appears to vary randomly across the surface~(see the supplementary material).
Nonetheless, through the optimization process, the beam could be directed to the desired direction, such that the main lobe intensity is increased by more than $6\,\mathrm{dB}$ in comparison to the initial voltage set. 
Note, however, that the voltages can only be optimized in one plane for the given prototype, such that the result is still well below the optimum. 
Hence, it can be stated that the beam is not ideally focused towards the desired direction, which is also the main reason for the comparably low aperture efficiency $\eta_\mathrm{Co}$ shown in Fig.~\ref{fig:Apt_eff_750}.

For a more detailed assessment of the impact of LC thickness on the measurement results, extended simulations are provided in the supplementary material.
The results reveal that a variation of the LC layer thickness has considerable impact on the beamforming performance of a \gls{RIS}. 
In particular, a high fluctuation across the surface poses a significant challenge.
Excess glue, non-uniform SU-8 photoresist spacers and dust particels are the main causes for the thickness variations.
A bias voltage optimization can help to mitigate the negative effects in laboratory environments, but may not be applicable in real world scenarios with \glspl{RIS} with thousands or millions of elements due to the excessive optimization time.
However, the achievement of uniform $t_\mathrm{LC}$ for large-scale \gls{RIS} is merely a technological challenge and already solved for \gls{LC} display applications.

\section{Comparison}
\begin{table*}[t!]
\centering
\captionsetup{justification=centering}
\caption{Comparison of \gls{LC}-\gls{RIS} realizations}
\begin{threeparttable}
\begin{tabular}{ccccccccccc}
\toprule
& Arch. & Year & $t_\mathrm{LC}$ & $f_\mathrm{0}$ & $BW$ & Nbr. El. & Aperture & Scan & $\Delta \phi_\mathrm{max}$ & Ref.\\ 
& & & (µm) & (GHz)  & (\%) &   & eff. (\%) & range &   \\ 
\midrule

1 & RA & 2023 & 200 & 30 & $\approx7$\tnote{a} & 12x12
& 33.4\tnote{a} 
& $\approx60\degree$ 
& $\approx240\degree$\tnote{a} 
& \cite{youn_liquid-crystal-driven_2023}\\

2 & RA\tnote{c} & 2023 & 250 & 28 & $\approx10$\tnote{a} & 10x10
& 22.1\tnote{a} 
& $\approx120\degree$  
& $\approx230\degree$\tnote{a} 
& \cite{kim_independently_2023}\\

3 & RA & 2024 & 200 & 38 & $\approx10$\tnote{a} & 12x12
& 22.9 
& $\approx35\degree$ 
& $\approx320\degree$\tnote{a} 
& \cite{li_development_2024}\\

4 & RA & 2024 & 12.5 & 275 & $\approx8$\tnote{a} & 32x32
& 
N/A 
& $\approx30\degree$\tnote{a} 
& $\approx300\degree$\tnote{a} 
& \cite{shen_liquid_2024}\\

5 & RA & 2024 & 50 & 415 & $\approx7$\tnote{a} & 32x32
& N/A 
& $\approx60\degree $
&
$\approx200\degree$\tnote{a} 
& \cite{xu_fully_2024}\\

6 & DLA & 2024 & 5 & 19.5 & $\approx8$\tnote{a} \,\tnote{b} & 48x48
& N/A 
& $\approx80\degree$\tnote{a}\ \tnote{b} 
& $>360\degree$\tnote{a} 
& \cite{wu_ultrathin_2024}\\

7 & RA & 2025 & 50 & 100 & $\approx3.5$\tnote{a} & 32x32
& N/A  
& $\approx100\degree$\tnote{a}\ \tnote{b} 
& $>360\degree$\tnote{a} 
& \cite{xu_liquid-crystal-based_2025}\\

8 &RA & 2025 & 50 & 345 & $\approx6$\tnote{a} & 32x32
& N/A 
& $\approx100\degree$\tnote{a}\ \tnote{b}
& $\approx280\degree$\tnote{a} 
& \cite{xu_reconfigurable_2025}\\

\vcentered{9} &
\vcentered{RA\tnote{c}\ \tnote{d}} &
\vcentered{2025} &
\vcentered{10} &
\vcentered{\shortstack{28.5 \\ 38}} &
\vcentered{\shortstack{$\approx11$\tnote{a} \\ $\approx8$\tnote{a}}} &
\vcentered{80×80} &
\vcentered{\shortstack{4.6\tnote{a} \\ 6\tnote{a}}} &
\vcentered{\shortstack{$\approx60^\circ$\tnote{a}\ \tnote{b} \\ $\approx60^\circ$\tnote{a}\ \tnote{b}}} &
\vcentered{\shortstack{$\approx240^\circ$\tnote{a} \\ $\approx240^\circ$\tnote{a}}} &
\vcentered{\cite{kim_millisecond-switching_2025}}
\\

\midrule

&  DLA & 2024 & 4.6 & 59 (61.9\tnote{e} \,) & $\approx24$ (9\tnote{e} \,) & 12x10
& 9.2 (\textit{17}\tnote{f} \,)  & $\approx110\degree$ & $>360\degree$ & This work \\

&  DLA & 2025 & 4.6 & 59 (60.5\tnote{e} \,) & $\approx24$ (9.5\tnote{e} \,)& 30x25
& 2.6 (\textit{16}\tnote{f} \,) & $\approx120\degree$ & $>360\degree$& This work \\

\bottomrule
\end{tabular}

\begin{tablenotes}
\small\itshape
\item[a] Estimated from graphs.
\item[b] Might be larger. Not more values provided.
\item[c] Dual polarized. Always minimum values shown regarding both polarizations.
\item[d] Combines resonant patches with resonant slots. Dual-band. 
\item[e] Bandwidth and center frequency stated with regard to the definition in this work (\SI{-3}{dB}) at $\theta_\mathrm{r} = 40\degree$. 
\item[f] Simulated. 
\end{tablenotes}
\end{threeparttable}
\label{Tab:RIS/SummaryOfAllPrototypes}
\end{table*}

To contextualize the results from this paper, the presented prototypes are compared to several recent reflect-type LC-RIS implementations from the literature in Table~\ref{Tab:RIS/SummaryOfAllPrototypes}. 
Note, however, that one or several specifications are often not directly provided in the literature, which is why they must be interpreted from graphs. 
When comparing the LC-RIS realizations, several observations can be made regarding the following key aspects.\\
\textbf{Type:}
The RA remains the most popular architecture for the design of LC-RIS, with only one prototype in the literature also employing a \gls{DLA} approach.\\
\textbf{LC layer thickness:}
To achieve few percents of bandwidth, \gls{RA}-based prototypes generally require thicker \gls{LC} layers~\cite{neuder_architecture_2024} than those utilizing the \gls{DLA}, leading to significantly slower response times. 
Nevertheless, the results in the table indicate that operating at higher frequencies enables the \gls{RA}-based prototypes to use thinner \gls{LC} layers, thanks to the shorter wavelengths. 
In contrast, the \gls{DLA} approaches feature thinner $\mathrm{t_\mathrm{LC}}$ also at lower frequencies, as the layer thickness has a much smaller impact on its performance when employing an appropriate phase shifter topology~\cite{neuder_architecture_2024}. 
An exception is prototype number 9, which is based on the \gls{RA}, but features $\mathrm{t_\mathrm{LC}}=\SI{10}{\mu \m}$ at \SI{28}{\GHz}.
This is accomplished by using patches placed on top of resonant cross-slot structures, enabling the thin $t_\mathrm{LC}$ while maintaining good functionality.\\
\textbf{Bandwidth:}
Before comparing the bandwidths, it must be noted that the latter is not stated directly in any of the literature sources.
Hence, due to the lack of data, it could not be estimated as the \SI{-3}{dB} bandwidth as defined in this paper.
The only metric for which the bandwidth could be compared is the variation of the differential \gls{RIS} induced phase shift over frequency.
For this purpose, the bandwidth is defined as the frequency range in which the differential phase shift, $\Delta \phi_\mathrm{max}$, does not differ more than 25\% from the value at its center frequency, that was defined in~\cite{Neuder_DualPol_2025}. 
However, it must be noted that this definition neglects the impact caused by the frequency dependent amplitude of the reflection, for example due to the limited bandwidth of the radiating element.

According to the utilized metric, the bandwidth of the prototypes introduced in this work is significantly higher than that of the prototypes found in the literature. 
This is primarily due to the choice of architecture.
Specifically, the \gls{RA}-based prototypes feature limited bandwidths because of the frequency dependence of their phase profile, which is characteristic to their resonant nature. 
In this context, prototype 6 is also expected to demonstrate a wide bandwidth, although the paper does not provide enough data to confirm this.\\
\textbf{Loss:}
Compared to the efficiencies of the \gls{RA}, the measured aperture efficiency of the prototypes presented in this work are considerably lower. 
Yet, two aspects must be noted:
First, with one exception, all realizations of the \gls{RA} type, for which the aperture efficiency can be determined, employ substantially thicker LC layers. 
With micrometer thin LC layers, their efficiency would drastically degrade.
Second, all the competing realizations use glass substrates with substantially lower loss (e.g. quartz glass, that is not suitable for \gls{RIS} with large apertures) and most use copper, which poses a significant difference to the materials used in this work, which are gold and comparably high loss glass (tan$\delta = 0.011$ versus tan$\delta = 0.0009-0.002$).

The only \gls{RA} approach with comparable \gls{LC} thickness (prototype 9) compared to the prototypes presented in this paper, though still more than twice as thick, likewise demonstrates a lower aperture efficiency than the small prototype. 
On top of that, the architecture and unit cell proposed in this work demonstrate the potential for high-efficiency \gls{LC}-\gls{RIS}, with simulations showing that efficiencies above 40\,\% are achievable when low-loss materials are utilized and significant fabrication tolerances are excluded.\\
\textbf{Scan and phase range:}
Most of the \gls{RA} implementations have limited phase ranges of less than 360°, which could reduce the beamforming capabilities of the LC-RIS. 
In contrast, the \gls{DLA} approaches can achieve a phase range exceeding 360°. 
The scan ranges are primarily determined by the radiating elements. 
While most of the competing implementations do not demonstrate beam steering at larger angles (> 40°), they are expected to be capable of steering towards these angles as well.

To conclude, while the conventional \gls{RA}-based prototypes do not allow for simultaneous optimization of loss, bandwidth and response time, prototype 9 emerges as the strongest \gls{RA}-based competitor.
By increasing the complexity of its design by introducing of resonant slots in its ground plane, it features a thin \gls{LC} layer and combines it with dual-band and dual-polarization functionality.
Still, the comparison highlights the superior potential of \gls{DLA} approaches, which promises micrometer-thin \gls{LC} layers, broad bandwidth, and the possibility of low-loss performance, in addition to exceptional versatility thanks to their wide phase range. 
The prototypes developed in this work reinforce these findings.
\section{Conclusion}
This paper presents the realization of two \SI{60}{\GHz} \gls{LC}-\gls{RIS} prototypes with 120 and 750 elements that are based on planar \gls{LC} phase shifters, i.e., the delay line architecture.
The \glspl{RIS} incorporate a thin $\mathrm{\SI{4.6}{\mu \m}}$ \gls{LC}-layer, demonstrating beam-steering capabilities of $\mathrm{\pm60\degree}$, \SI{-3}{\dB} bandwidths exceeding 9\% a measured aperture efficiency of more than 9\%. 
While fabrication constraints impact the performance, this work confirms \\
i) the scalability of the \gls{LC}-\gls{RIS}, as such challenges have already been effectively addressed in large-scale \gls{LC}-display manufacturing.\\
ii) the superior potential of designing \gls{LC}-\gls{RIS} based on the delay line architecture in contrast to the conventional resonant architecture, particularly regarding the trade-off between bandwidth, efficiency, and response time.



\section*{Acknowledgment}
This work was funded by the Deutsche Forschungsgemeinschaft (DFG, German Research Foundation) - Project-ID 287022738 - TRR 196 MARIE within project C09. In addition, thanks goes to Merck Electronics KGaA, Darmstadt, Germany, for providing the liquid crystal mixture.


\section*{Competing interests}
Competing interests: The authors declare none.

\printbibliography


\end{document}